\documentclass{desyproc}

\begin{document}
\title{HERA Inclusive Diffraction \& Factorisation Tests}

\author{{\slshape Paul Newman (for the H1 and ZEUS Collaborations)}\\[1ex]
School of Physics \& Astronomy, University of Birmingham, B15 2TT, UK}

\contribID{38}

\confID{1407}  
\desyproc{DESY-PROC-2009-03}
\acronym{PHOTON09} 
\doi  

\maketitle

\begin{abstract}
HERA measurements of diffractive $ep$ scattering - 
the quasi-elastic scattering of the photon
in the proton colour field - are summarised. 
Emphasis is placed on the most recent data. 
\end{abstract}

\section{Introduction}

Between 1992 and 2007, the 
HERA accelerator provided $ep$ collisions at centre
of mass energies beyond $300 \ {\rm GeV}$
at the interaction points of the H1 and ZEUS experiments. 
Perhaps the most interesting results to emerge relate to
the newly accessed field of 
perturbative strong interaction physics at low Bjorken-$x$,
where parton densities become extremely large \cite{max}.
Questions arise as to how and where non-linear dynamics tame
the parton density growth \cite{saturation} 
and challenging features such as geometric scaling \cite{geo:scale}
are observed. Central to this low $x$ physics landscape 
is a high rate of diffractive
processes, in which a colourless exchange takes place and
the proton remains intact. 
In particular, the study of semi-inclusive diffractive
deep-inelastic scattering (DDIS),
$\gamma^* p \rightarrow X p$ \cite{hera:diff} 
has led to a revolution in our microscopic, parton level,
understanding of the structure of elastic
and quasi-elastic high energy hadronic scattering \cite{hebecker:review}. 
Comparisons
with hard diffraction in proton-(anti)proton scattering have
also improved our knowledge of absorptive 
and underlying event effects in which the 
diffractive signature may be obscured by multiple interactions in the
same event \cite{gap:survival}. In addition to their fundamental
interest in their own right, these issues are highly relevant
to the modelling of chromodynamics at the LHC \cite{lhcdiff}.  
 
\begin{figure}[h]
\centerline{\hspace*{0.1cm}
            {\Large{\bf{(a)}}}
            \includegraphics[height=0.3\textwidth]{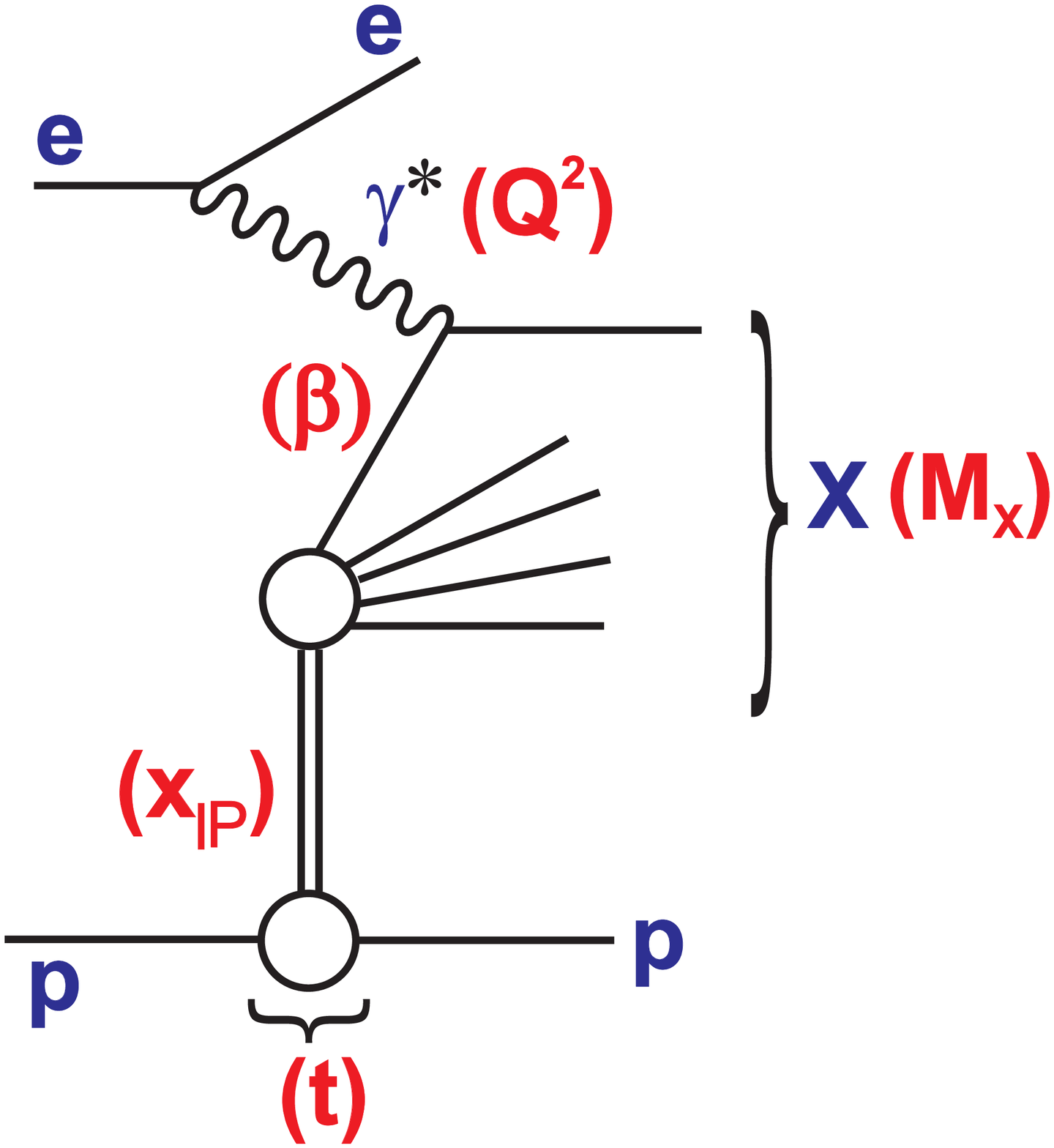}
            \hspace*{0.1cm}
            {\Large{\bf{(b)}}}
            \includegraphics[height=0.3\textwidth]{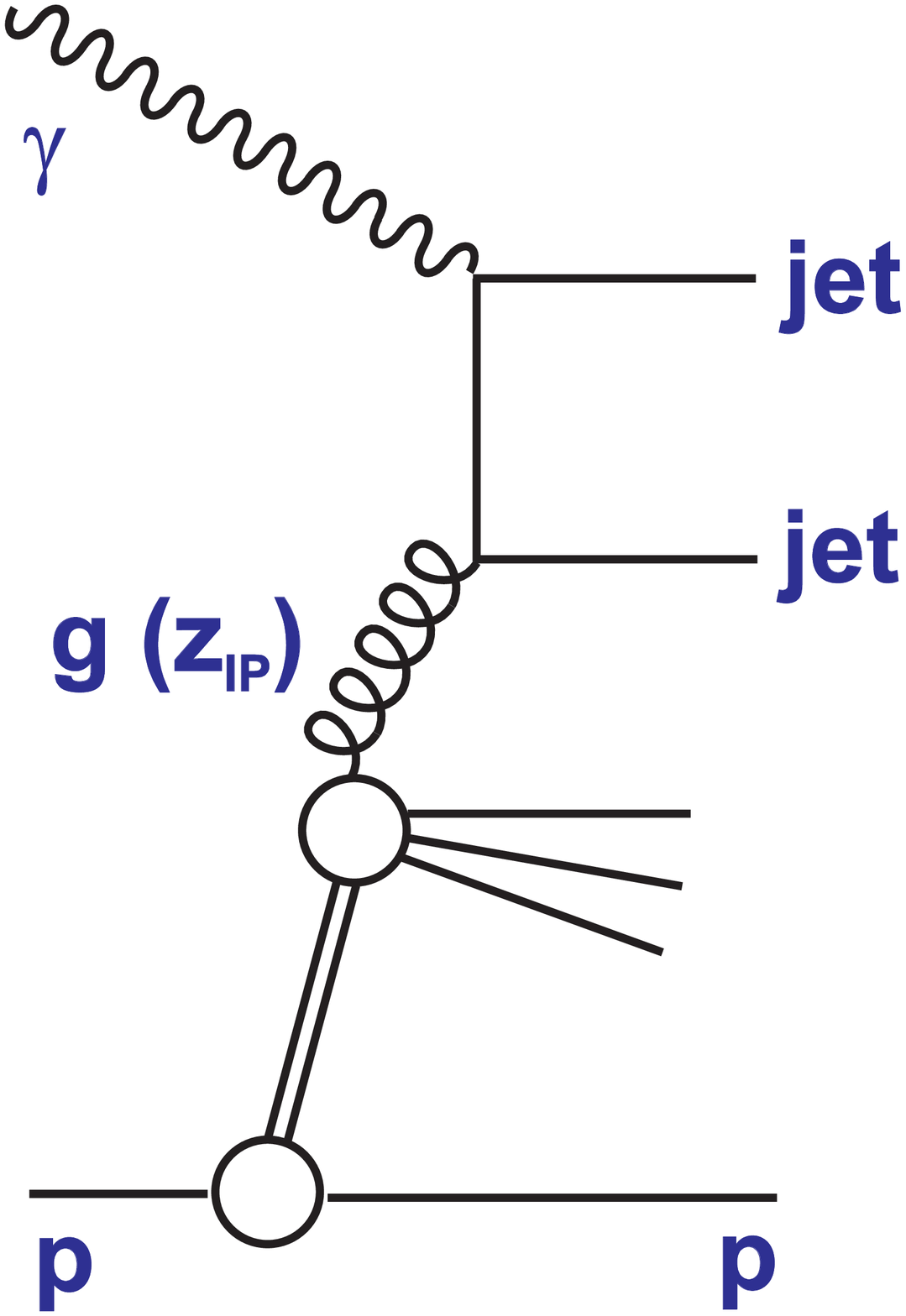}
            \hspace*{0.6cm}
            {\Large{\bf{(c)}}}
            \includegraphics[height=0.3\textwidth]{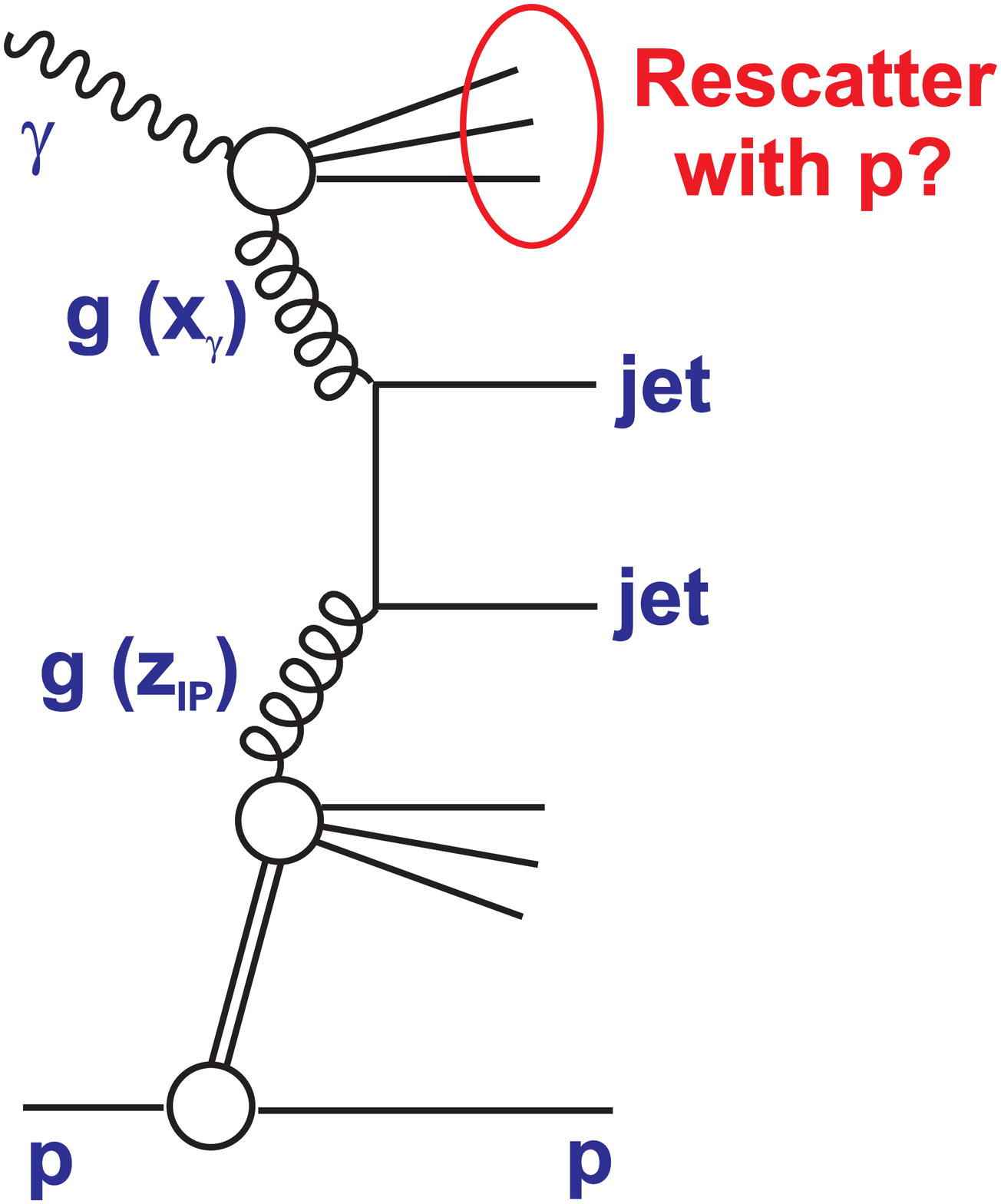}}
\caption{Sketches of diffractive $ep$ processes.
(a) Inclusive DDIS at the level of the quark parton model,
illustrating the kinematic variables 
discussed in the text. (b) Dominant leading order diagram for 
hard scattering in DDIS or direct photoproduction, in
which a parton of momentum fraction $z_{I\!\!P}$ from the DPDFs
enters the hard scattering. (c) A leading
order process in resolved photoproduction involving a parton 
of momentum fraction $x_\gamma$ relative to the photon.}
\label{feynman}
\end{figure}

The kinematic variables describing 
DDIS are illustrated in figure~\ref{feynman}a.
The longitudinal momentum fractions
of the colourless exchange with respect to the incoming
proton and of the struck quark with respect to the colourless
exchange are denoted $x_{_{I\!\!P}}$ and $\beta$, respectively,
such that $\beta \, x_{_{I\!\!P}} = x$.
The squared four-momentum transferred at
the proton vertex is given by the Mandelstam $t$ variable.
The semi-inclusive DDIS cross section is usually presented in 
the form of a diffractive reduced cross section $\sigma_r^{D(3)}$,
integrated over $t$ and related to the experimentally measured
differential cross section by \cite{h1:lrg}
\begin{equation}
\frac{{\rm d}^3\sigma^{ep \rightarrow e X p}}{\mathrm{d} x_{_{I\!\!P}} \ \mathrm{d} x \ \mathrm{d} Q^2} = \frac{2\pi
  \alpha^2}{x Q^4} \cdot Y_+ \cdot \sigma_{r}^{D(3)}(x_{_{I\!\!P}},x,Q^2) \ ,
\label{sigmar}
\end{equation}
where $Y_+ = 1 + (1-y)^2$ and $y$ is the usual Bjorken variable.
The reduced cross section depends 
at moderate scales, $Q^2$, on two
diffractive structure functions
$F_2^{D(3)}$ and $F_L^{D(3)}$ according to
\begin{equation}
\sigma_r^{D(3)} =
F_2^{D(3)} - \frac{y^2}{Y_+} F_L^{D(3)}.
\label{sfdef}
\end{equation}
For $y$ not too close to unity,
$\sigma_r^{D(3)} = F_2^{D(3)}$ holds to very good approximation.

\section{Measurement methods and comparisons}

Experimentally, diffractive $ep$ scattering
is characterised by the presence of a leading proton in the
final state, retaining most of the initial state proton energy, and
by a lack of hadronic activity in the
forward (outgoing proton) direction, such that the
system $X$ is cleanly separated and 
its mass $M_X$ may be measured in the central
detector components.
These signatures have been widely exploited at HERA to select
diffractive events by tagging the outgoing proton
in the H1 Forward Proton Spectrometer or
the ZEUS Leading Proton Spectrometer 
(`LPS method' \cite{h1:fps,zeus:lrglps,h1:fpshera2}) or
by requiring the presence of a large gap in the rapidity
distribution of hadronic final state particles
in the forward region 
(`LRG method' \cite{h1:lrg,zeus:lrglps,H1:newdata}).
In a third approach, not considered in detail here, 
the inclusive DIS sample is
decomposed into diffractive and non-diffractive contributions based
on their characteristic dependences on $M_X$ \cite{H1:newdata,zeus:mx}.
Whilst the LRG and $M_X$-based techniques yield better
statistics than the LPS method, they suffer from systematic uncertainties 
associated with an admixture 
of proton dissociation to low mass states, which 
is irreducible due to the limited forward detector acceptance.

\begin{figure}[h]
\centerline{\includegraphics[width=0.4\textwidth]{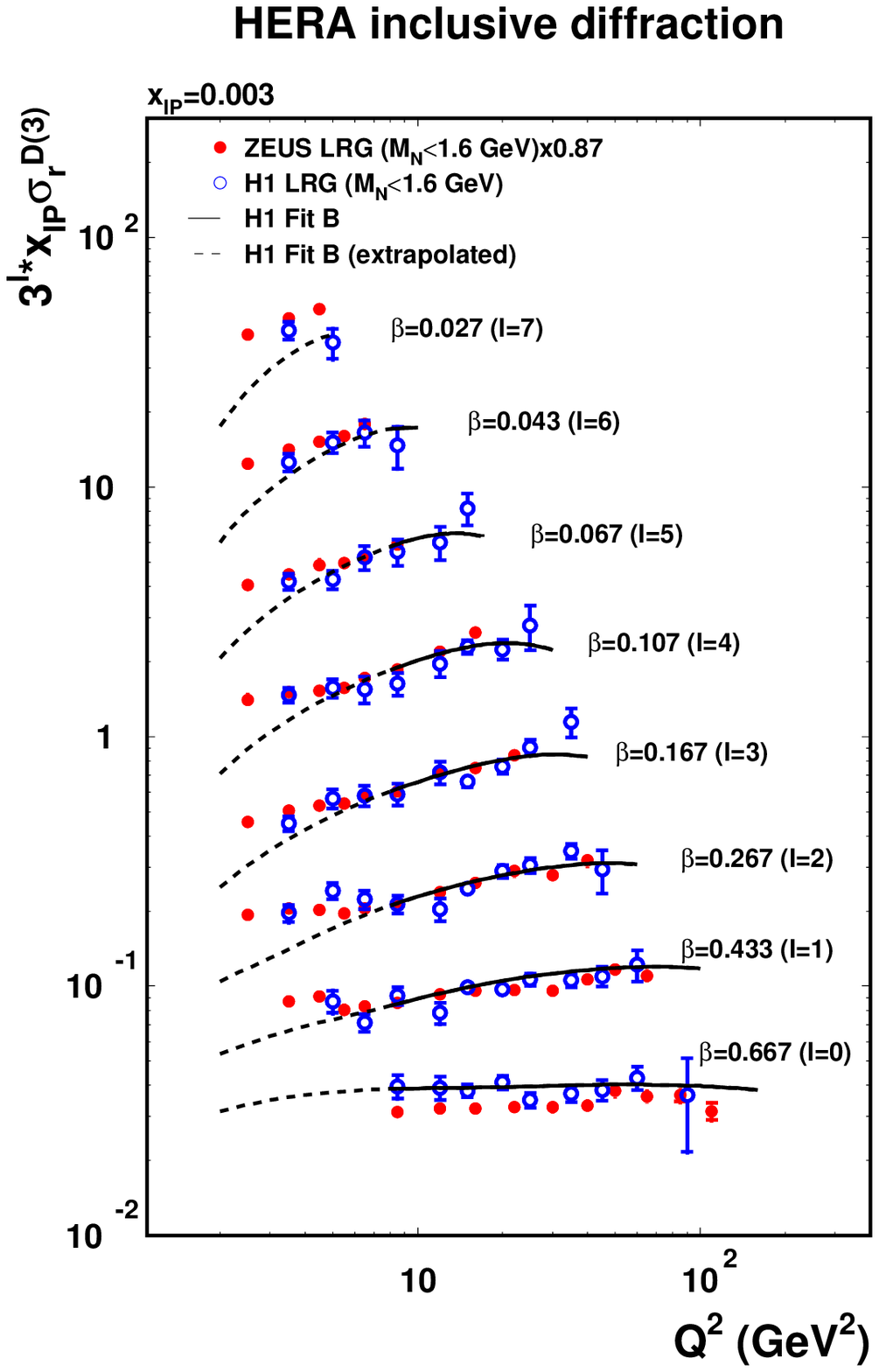}
            \hspace*{1cm}
            \includegraphics[width=0.4\textwidth]{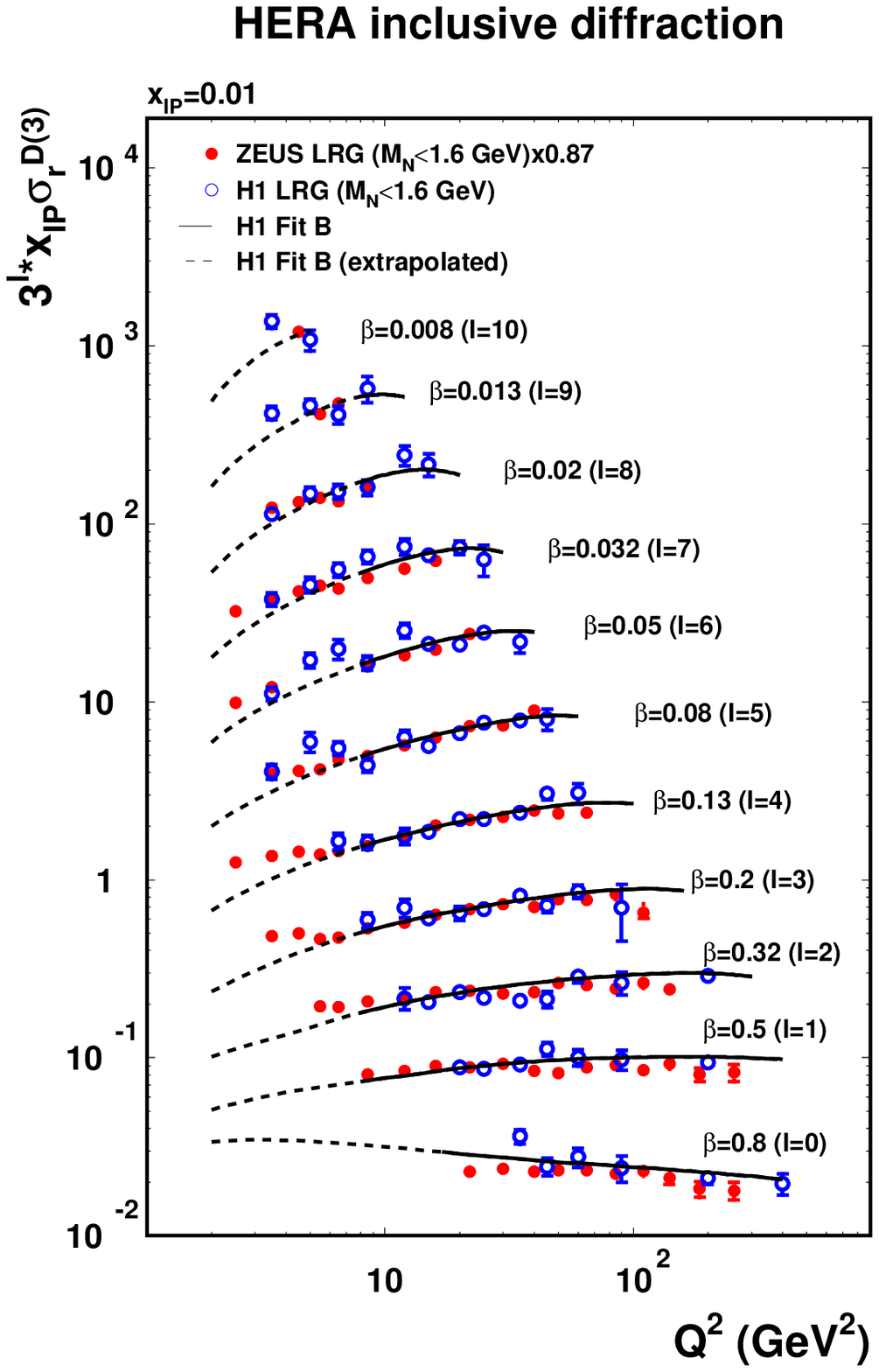}}
\caption{H1 and ZEUS measurements of the diffractive reduced cross section
at two example $x_{I\!\!P}$ values \cite{newman:ruspa}. 
The ZEUS data are
scaled by a factor of $0.87$ to match the H1 normalisation. 
The data are compared with the results of the  
H1 2006 Fit B DPDF based parameterisation \cite{h1:lrg}
for $Q^2 \geq 8.5 \ {\rm GeV^2}$
and with its DGLAP based extrapolation to lower $Q^2$.}
\label{h1vzeus}
\end{figure}

The H1 collaboration recently released a preliminary 
proton-tagged measurement
using its full available FPS sample at HERA-II \cite{h1:fpshera2}. 
The integrated luminosity
is $156 \ {\rm pb^{-1}}$, a factor of 20 beyond previous H1
measurements. The new data tend to lie slightly above
the recently published 
final ZEUS LPS data from HERA-I \cite{zeus:lrglps}, but are
within the combined normalisation uncertainty of around $10\%$.
The most precise test of compatibility between H1 and ZEUS is
obtained from the LRG data. The recently published ZEUS data \cite{zeus:lrglps}
are based on an integrated luminosity of $62 \ {\rm pb^{-1}}$ and
thus have substantially improved statistical precision compared
with the older H1 published results \cite{h1:lrg}. The normalisation 
differences between the two experiments are most obvious here,
having been quantified at $13\%$, which is a little beyond
one standard deviation in the combined normalisation uncertainty.
After correcting for this factor, very good agreement is observed
between the shapes of the H1 and ZEUS cross sections throughout most of
the phase space studied, as shown in 
figure~\ref{h1vzeus}.
A more detailed comparison between different diffractive cross section
measurements by H1 and ZEUS and a first attempt to combine the 
results of the two experiments can be found in \cite{newman:ruspa}.

\section{Soft physics at the proton vertex}
\label{soft}

\begin{figure}[h]
\centerline{\hspace*{0.3cm}
            {\Large{\bf{(a)}}}
            \includegraphics[width=0.45\textwidth]{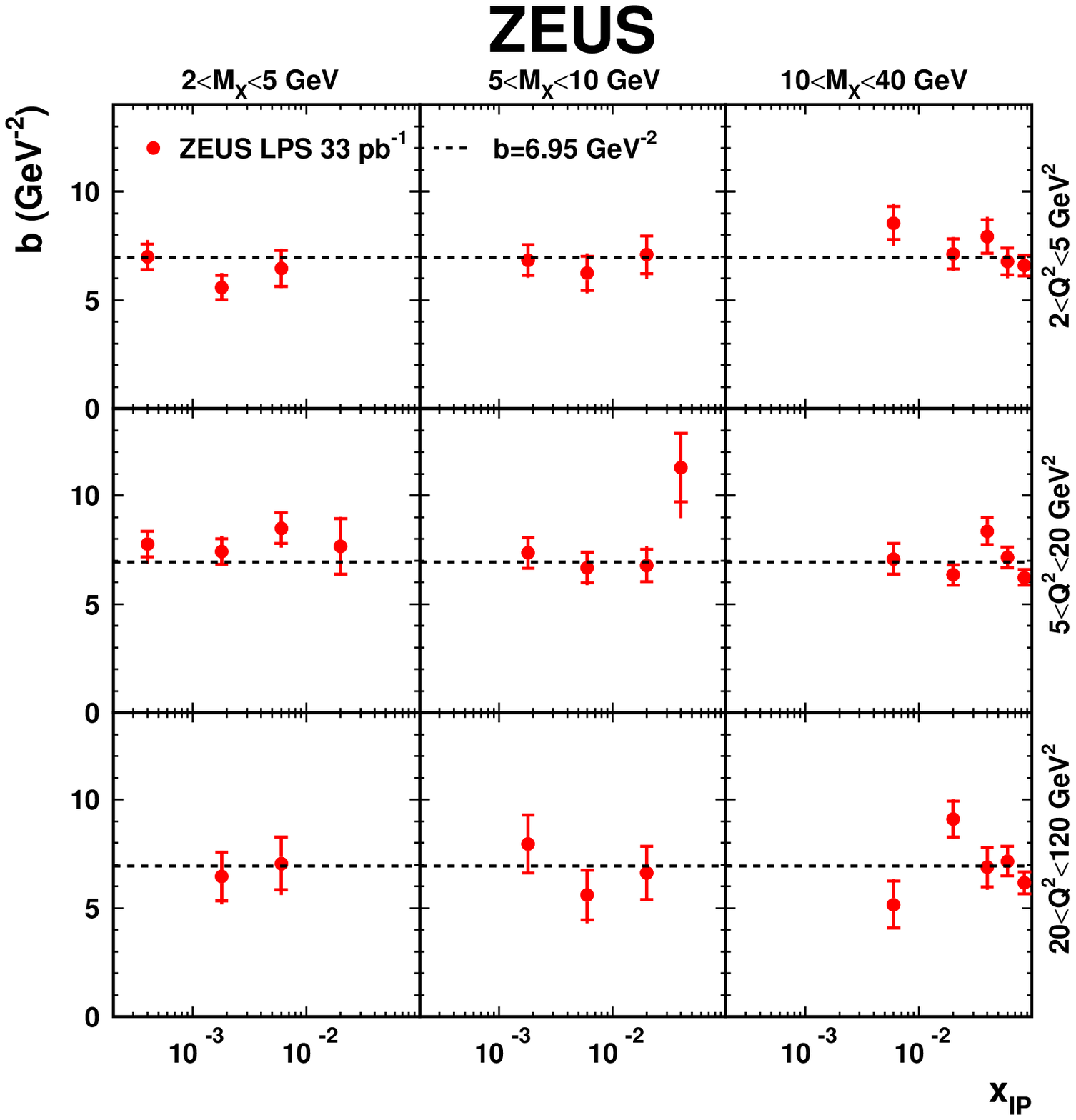}
            \hspace*{0.3cm}
            {\Large{\bf{(b)}}}
            \hspace*{-0.3cm}
            \includegraphics[width=0.45\textwidth]{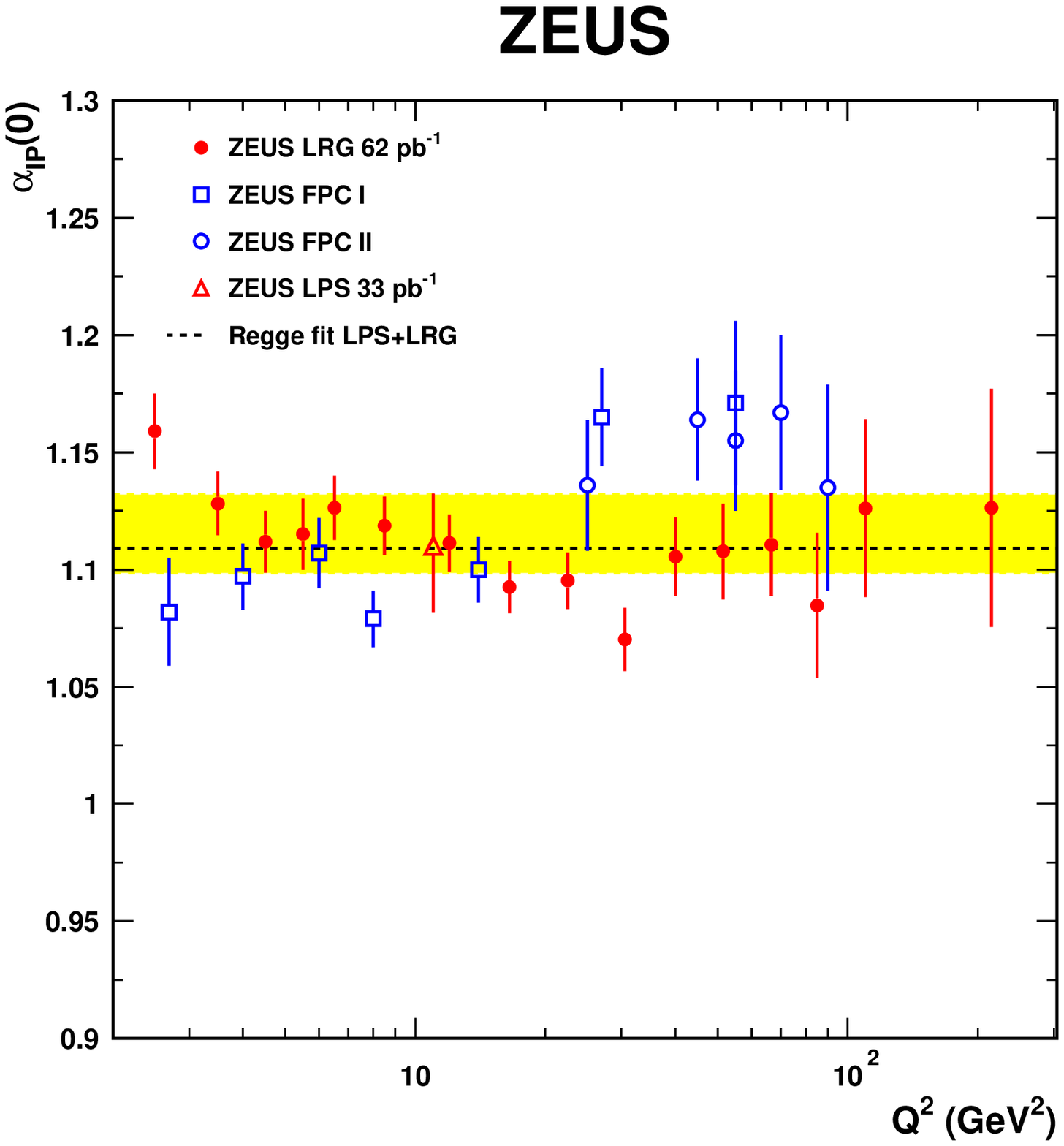}}
\caption{a) Measurements of the exponential $t$ slope from ZEUS
LPS data, shown as a function of $Q^2$, $x_{I\!\!P}$ and $M_X$.
b) ZEUS extractions of the effective pomeron intercept describing
the $x_{I\!\!P}$ dependence of DDIS data at different $Q^2$ 
values \cite{zeus:lrglps}.}
\label{regge}
\end{figure}

To good approximation, LRG and LPS 
data show \cite{h1:lrg,h1:fps,zeus:lrglps} that 
DDIS data satisfy a `proton vertex 
factorisation',\footnote{This factorisation is not expected to hold 
to indefinite precision, due for example to the presence of a `hard',
fully perturbatively tractable diffractive exchange which governs
exclusive
vector meson production in the presence of hard scales \cite{vms}.
This leads to a higher twist contribution to $\sigma_r^D$ for 
$\beta \rightarrow 1$ \cite{bekw,hebecker:teubner}. However, 
this contribution seems to be numerically small when compared with the the
inclusive diffractive cross section.}
whereby 
the dependences on variables which describe the scattered 
proton ($x_{I\!\!P}$, $t$) factorise from those
describing the hard partonic interaction ($Q^2$, $\beta$).
For example, 
the slope parameter $b$, extracted in \cite{zeus:lrglps}
by fitting the $t$ distribution to the form
${\rm d} \sigma / {\rm d} t \propto e^{b t}$, is shown as a function
of DDIS kinematic variables in figure~\ref{regge}a. 
There are no significant variations
from the average value of $b \simeq 7 \ {\rm GeV^{-2}}$ anywhere in the
studied range.   
The measured value of $b$ is significantly larger than
that from `hard' 
exclusive vector meson production ($ep \rightarrow eVp$) \cite{vms}. 
It is 
characteristic of an interaction region 
of spatial extent considerably larger than the proton radius, 
indicating that the dominant feature of DDIS is the probing with
the virtual photon of
non-perturbative exchanges similar to the pomeron 
of soft hadronic physics \cite{pomeron}. 

Figure~\ref{regge}b shows the $Q^2$ dependence of the effective
pomeron intercept $\alpha_{I\!\!P} (0)$, which is extracted from 
the $x_{I\!\!P}$ dependence of the data \cite{zeus:lrglps}. 
No significant
dependence on $Q^2$ is observed, again compatible with
proton vertex factorisation. 
These results are consistent with the H1 value of
$\alpha_{I\!\!P}(0) = 1.118 \pm 0.008 \ {\rm (exp.)} \
^{+0.029}_{-0.010} \ {\rm (model)}$ \cite{h1:lrg}.
Both collaborations have also extracted a value for the slope of
the effective pomeron trajectory, the recently published ZEUS
value being 
$\alpha_{I\!\!P}^\prime = -0.01 \pm 0.06 \ {\rm (stat.)}
\ \pm 0.06 \ {\rm (syst.)} \ {\rm GeV^{-2}}$ \cite{zeus:lrglps}.

The intercept 
of the effective pomeron trajectory
is consistent within errors with the `soft pomeron' results 
from fits to total cross sections and
soft diffractive data \cite{dl}.
Although larger effective intercepts 
have been measured in hard vector meson production \cite{vms}, 
no deviations with either $Q^2$ or $\beta$ have yet been observed
in inclusive DDIS.   
The measured slope of the effective trajectory is smaller than the canonical
soft diffractive value of $0.25 \ {\rm GeV^{-2}}$ \cite{soft:alphaprime},
though it is compatible with results from the soft 
exclusive photoproduction of $\rho^0$ mesons at HERA \cite{blist}.

\section{Diffractive Parton Density Functions}
\label{sec:dpdfs}

\begin{figure}[h]
\centerline{\includegraphics[width=0.4\textwidth]{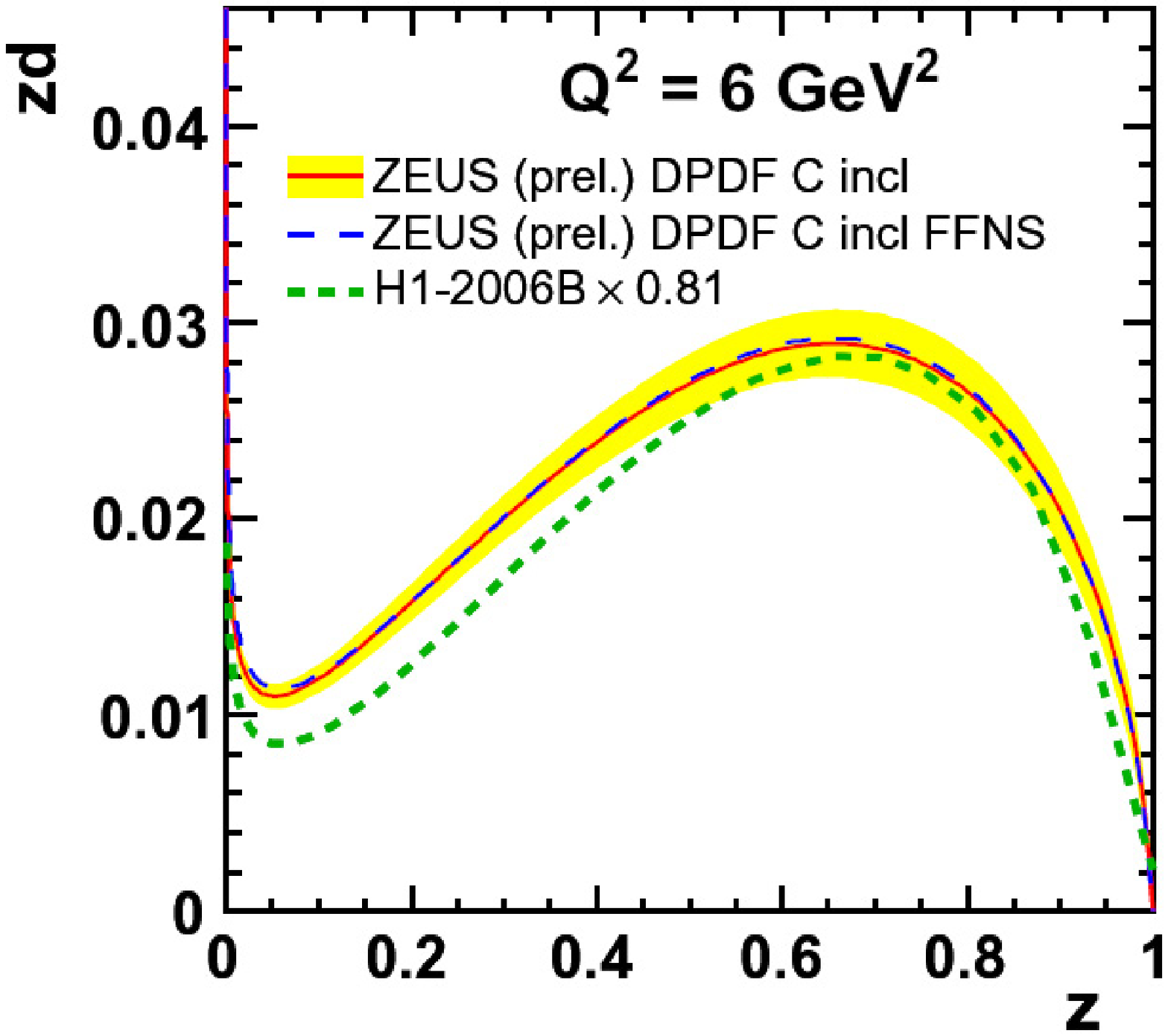}
            \hspace*{1cm}
            \includegraphics[width=0.3875\textwidth]{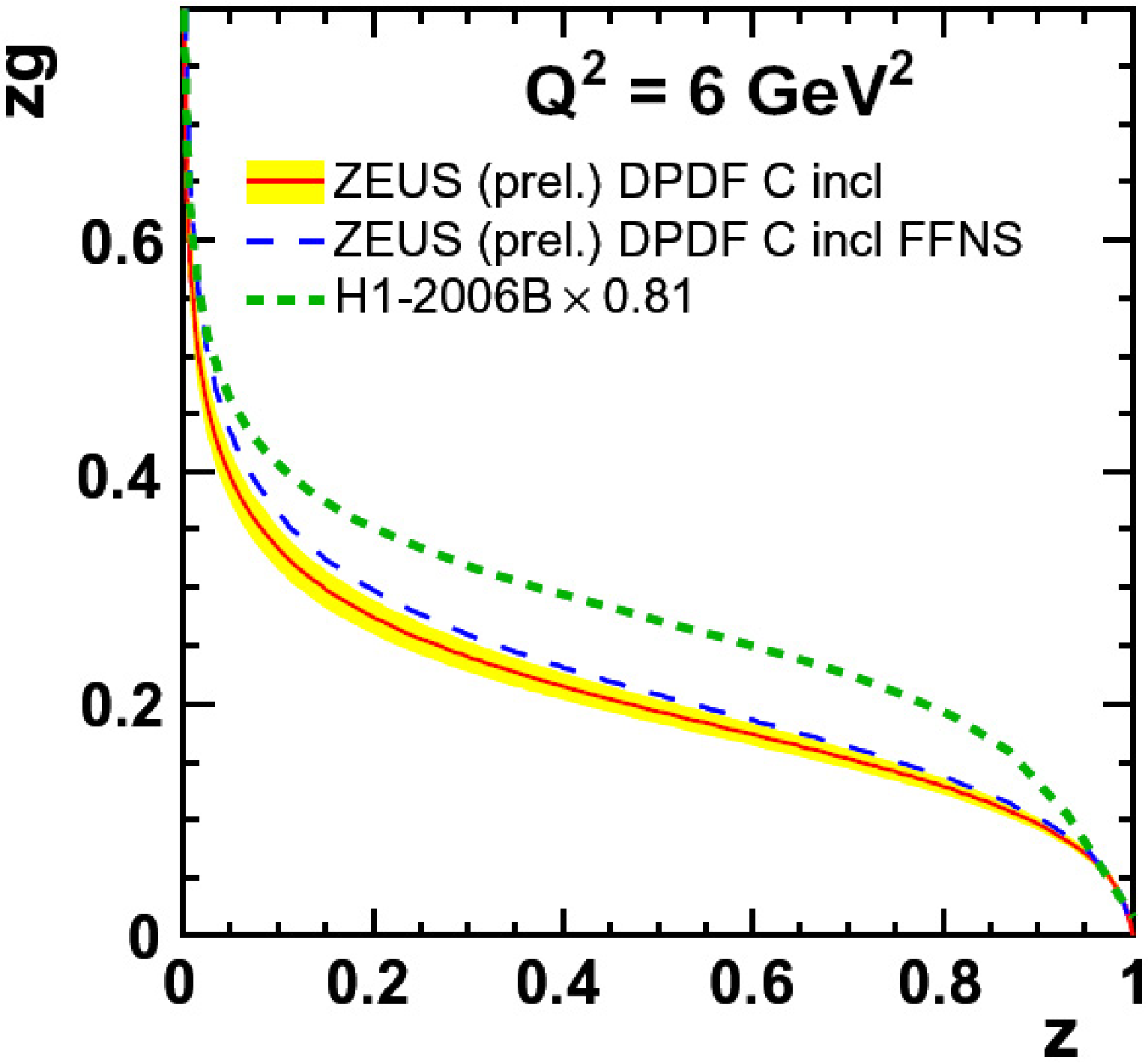}}
\caption{ZEUS down quark (one sixth of the total quark $+$ 
antiquark) and gluon densities
as a function of generalised momentum fraction $z$ at 
$Q^2 = 6 \ {\rm GeV^2}$ \cite{zeus:diffqcd}. 
Two heavy flavour schemes are shown, 
as well as H1 results \cite{h1:lrg} corrected
for proton dissociation with a factor of 0.81.}
\label{dpdfs}
\end{figure}

In the framework of the proof \cite{collins} of a
hard scattering collinear QCD factorisation theorem
for semi-inclusive DIS processes such as DDIS, the concept of `diffractive
parton distribution functions' (DPDFs)~\cite{facold}
may be introduced, representing conditional proton parton probability
distributions under the constraint of a leading final state proton with a
particular four-momentum.
The differential DDIS cross 
section may then be written in terms of convolutions of
partonic cross sections $\hat{\sigma}^{e i} (x, Q^2)$ with
DPDFs $f_i^D$ as
\begin{equation}
{\rm d} \sigma^{ep \rightarrow eXp} (x, Q^2, x_{_{I\!\!P}}, t) = \sum_i \
f_i^D(x, Q^2, x_{_{I\!\!P}}, t) \ \otimes \
{\rm d} \hat{\sigma}^{ei}(x,Q^2) \ .
\label{equ:diffpdf}
\end{equation}
The empirically motivated proton vertex factorisation 
property (section~\ref{soft}) suggests a 
further factorisation, whereby the DPDFs vary only
in normalisation with the four-momentum of the final state proton
as described by $x_{_{I\!\!P}}$ and $t$:
\begin{equation}
f_i^D(x,Q^2,x_{_{I\!\!P}},t) = f_{I\!\!P/p}(x_{_{I\!\!P}},t) \cdot
f_i (\beta=x/x_{_{I\!\!P}},Q^2) \ .
\label{reggefac}
\end{equation}
Parameterising $f_{I\!\!P/p}(x_{_{I\!\!P}},t)$
using Regge asymptotics, equation~\ref{reggefac}
amounts to a description of
DDIS in terms of the exchange of a factorisable
pomeron with universal parton densities \cite{ingelman:schlein}.
The 
$\beta$ and $Q^2$ dependences
of $\sigma_r^D$ may then be subjected to a perturbative QCD 
analysis based on the DGLAP equations in order to obtain DPDFs.
Whilst $F_2^D$ directly measures
the quark density, the gluon density is only indirectly
constrained, via the scaling violations 
$\partial F_2^D / \partial \ln Q^2$.

The high statistics ZEUS LRG and LPS data \cite{zeus:lrglps}
have recently been fitted to extract DPDFs \cite{zeus:diffqcd}.
The method and DPDF parameterisation are similar to an earlier
H1 analysis \cite{h1:lrg}, the main step forward being in the heavy 
flavour treatment,
which now follows the general mass variable flavour number 
scheme \cite{gm:vfns}.
In figure~\ref{dpdfs},
the resulting DPDFs are compared with results from both ZEUS and H1 
using a fixed flavour number scheme. 
The agreement between the experiments
is reasonable when the uncertainty on the H1 DPDFs is also taken into
account and the conclusion that the dominant feature is a
gluon density with a relatively hard $z$ dependence is confirmed. 
The error bands shown 
in figure~\ref{dpdfs} represent experimental uncertainties only.
Whilst the quark densities are rather well known throughout the
phase space, the theoretical uncertainties
on the gluon density are large. Indeed, in the 
large $z$ region, where the 
dominant parton splitting is $q \rightarrow qg$, the
sensitivity of $\partial F_2^D / \partial \ln Q^2$ 
to the gluon density becomes poor
and different DPDF parameterisations 
lead to large variations 
\cite{h1:lrg,zeus:diffqcd}. Improved large $z$ constraints 
have been
obtained by including dijet data in the QCD 
fits \cite{matthias,zeus:diffqcd}.

In common with the inclusive proton
PDFs at low $x$ \cite{max}, the DPDFs exhibit a
ratio of around 7:3 between gluons and quarks, consistent
with a common QCD radiation pattern far from the valence region. 
Qualitatively, the diffractive quark density is similar
in shape to that of the photon \cite{photon}, which might be expected if
the high $z$ quarks are generated from initial basic
$g \rightarrow q \bar{q}$ splittings, similar to the 
$\gamma \rightarrow q \bar{q}$ splitting in the photon case.

\section{Factorisation Tests in Diffractive DIS}

\begin{figure}[h]
\centerline{\includegraphics[width=0.4\textwidth]{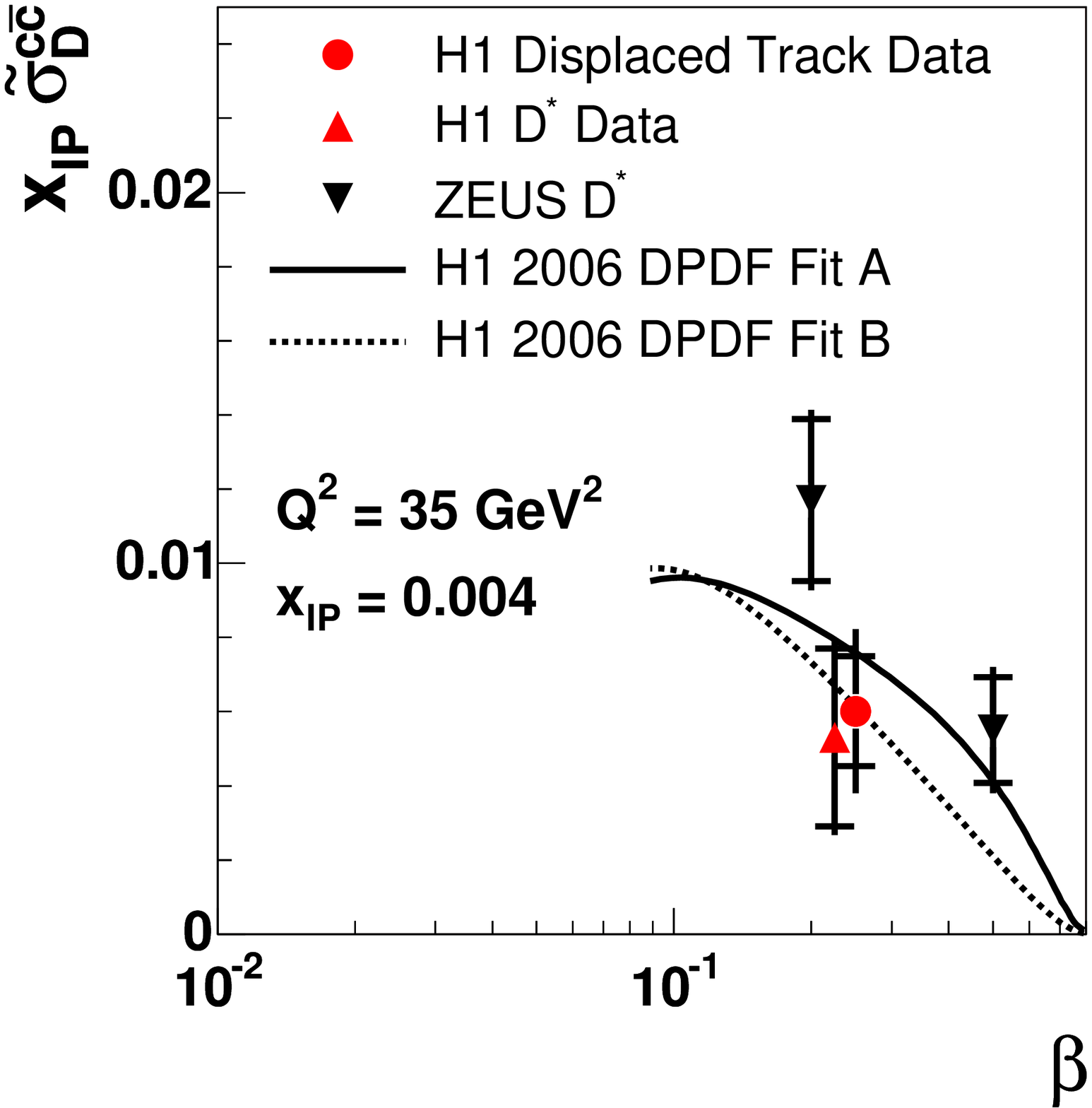}
            \hspace*{1cm}
            \includegraphics[width=0.4\textwidth]{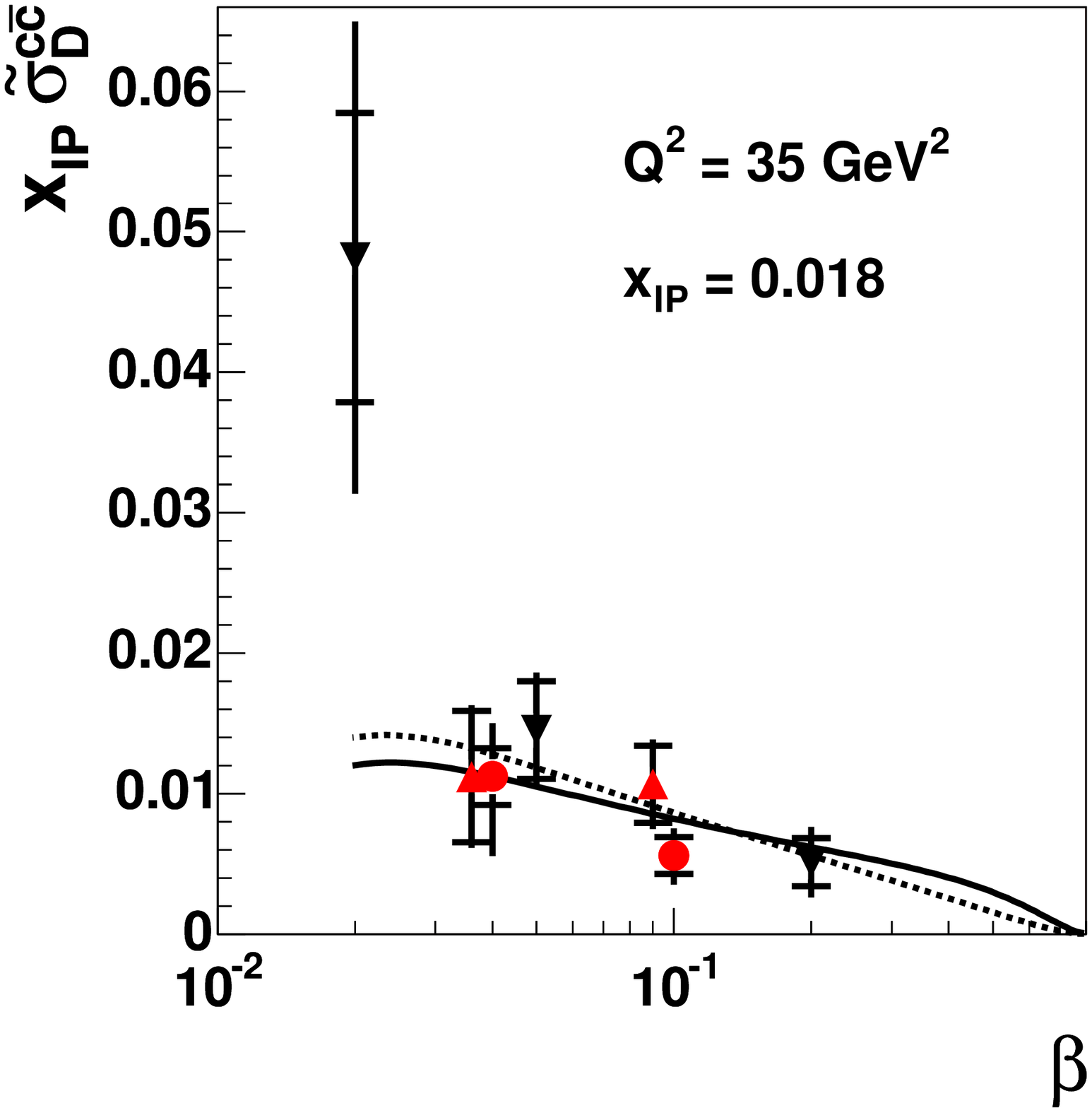}}
\vspace*{-0.3cm}
\caption{Comparisons \cite{pault}
of measurements of diffractive open charm
production with predictions based on DPDFs extracted from
$\sigma_r^D$ data \cite{h1:lrg}.}
\label{charm}
\end{figure}

According to \cite{collins}, the diffractive parton densities
extracted from $\sigma_r^D$ should be applicable to the prediction
of a wide range of other observables in diffractive DIS. There 
have been many tests of this diffractive hard scattering factorisation
over the years, the most precise and
detailed arising from jet \cite{dis:jets,sebastian,matthias}
and heavy flavour \cite{dis:charm,pault} cross section measurements. 
Being dominated by the boson-gluon fusion parton level
process $\gamma^* g \rightarrow q \bar{q}$ (figure~\ref{feynman}b), these
data are directly sensitive 
to the diffractive gluon density, in contrast to $\sigma_r^D$.
Such tests have been successful at moderate values of
$z$, as shown for the example of diffractive charm quark 
production in figure~\ref{charm}. 
As mentioned in section~\ref{sec:dpdfs},
the situation changes at large 
$z \, {\raisebox{-0.7ex}{$\stackrel {\textstyle>}{\sim}$}} \, 0.4$,
where the gluon density from $\sigma_r^D$ 
has a large uncertainty and dijet data 
give the best constraints.

\begin{wrapfigure}{r}{0.47\columnwidth}
\includegraphics[width=0.45\columnwidth]{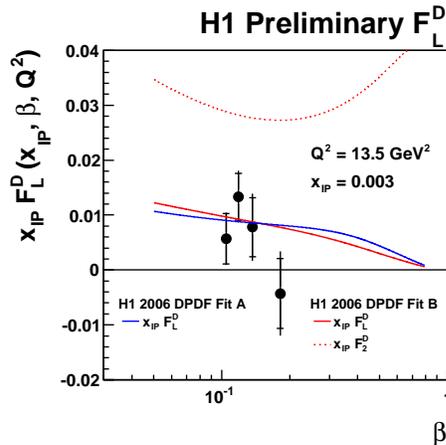}
\vspace*{-0.7cm}
\caption{First measurement of the longitudinal diffractive structure
function \cite{h1:fld}, compared with DPDF based predictions.}
\label{fl}
\end{wrapfigure}

At low $x$ and $Q^2$, the
longitudinal diffractive structure function,
$F_L^D$, is closely related to the diffractive
gluon density \cite{prn:fld} and thus gives a complementary test of
diffractive factorisation and the role of gluons to those provided 
by jet and charm cross sections. 
Measurements of $F_L^D$ became
possible following the reduced proton beam energy runs at the end of
HERA operation. According to equation~\ref{sfdef}, $F_L^D$ and $F_2^D$
may then be separated through the $y = Q^2 / (s \, \beta \, x_{_{I\!\!P}})$ 
dependence as $s$ varies at fixed $Q^2$, $\beta$ and $x_{_{I\!\!P}}$. 

The H1 collaboration recently
released preliminary $F_L^D$ data, as shown in figure~\ref{fl}.
The results \cite{h1:fld}, 
when integrated over $\beta$ show that $F_L^D$ is non-zero
at the $3 \sigma$ level. It is also clearly incompatible with its maximum
possible value of $F_2^D$. The measured ratio of longitudinal
to transverse photon induced cross sections in diffraction is similar
to that in inclusive DIS measurements \cite{fl:inclusive}, though
the errors in the diffractive case are large. 
The measured $F_L^D$
is in agreement with all reasonable 
predictions based on DPDFs
extracted from $\sigma_r^D$. Dipole model predictions such as 
\cite{saturation,bekw} 
have thus far neglected any contribution from a leading twist
$F_L^D$ and formally give predictions very close to zero in the 
relatively low $\beta$ range covered. A hybrid approach \cite{kgb:a}
which mixes
a leading twist DPDF based $F_L^D$ with a higher twist contribution
at high $\beta$ derived from \cite{saturation}, is in good agreement 
with the data.  

\section{Hard Diffractive Photoproduction and Rapidity Gap Survival 
Probabilities}

\begin{wrapfigure}{r}{0.45\columnwidth}
\includegraphics[width=0.45\columnwidth]{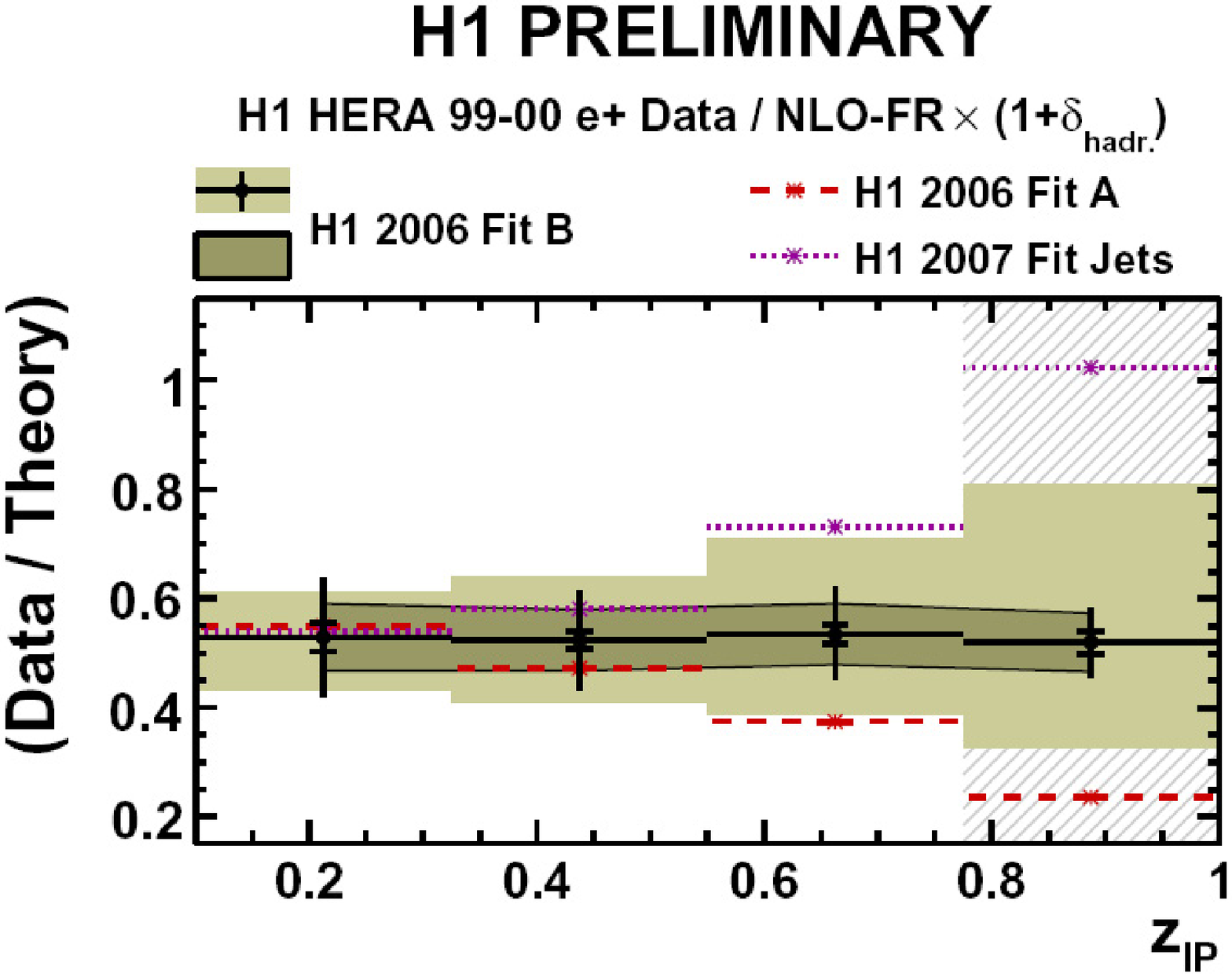}
\includegraphics[width=0.45\columnwidth]{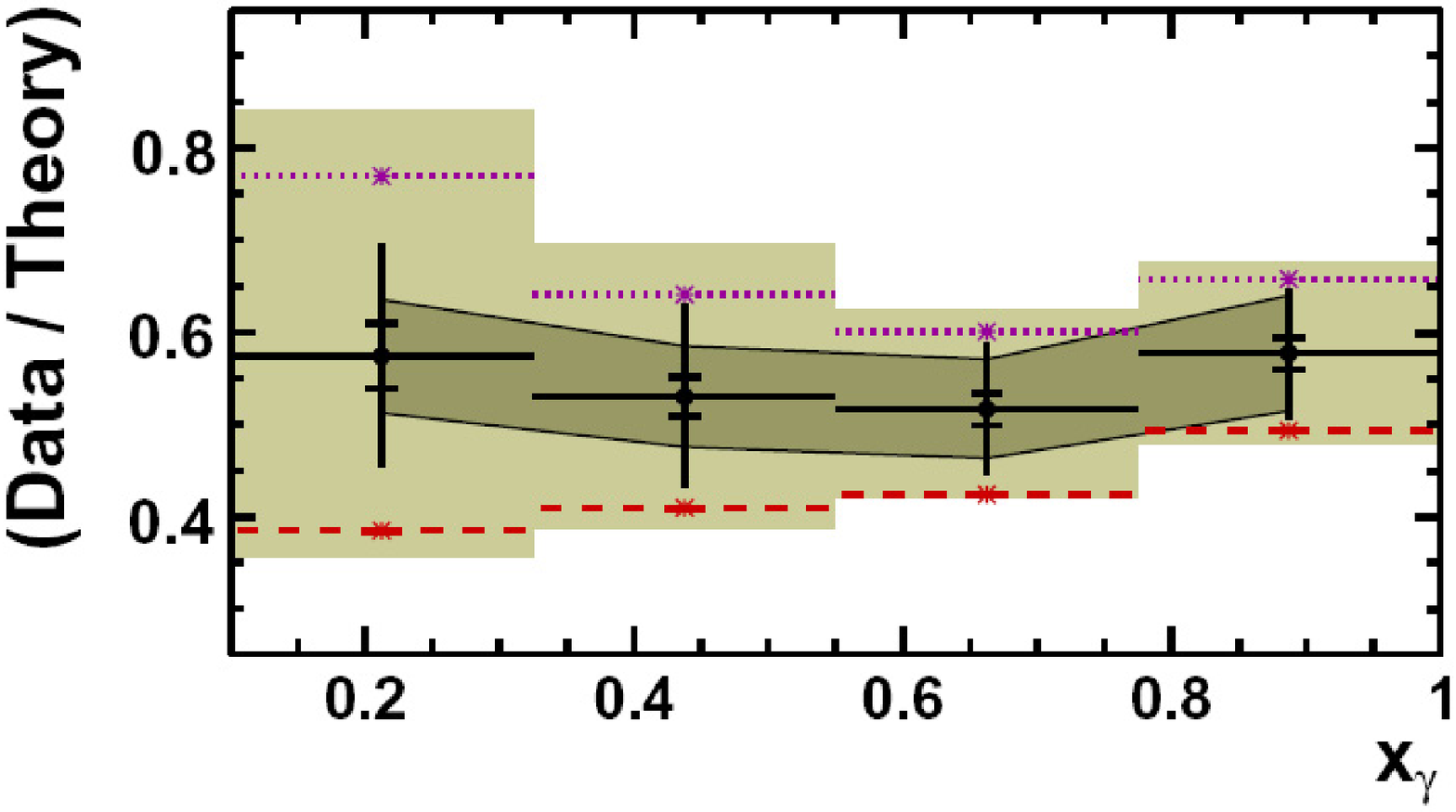}
\includegraphics[width=0.45\columnwidth]{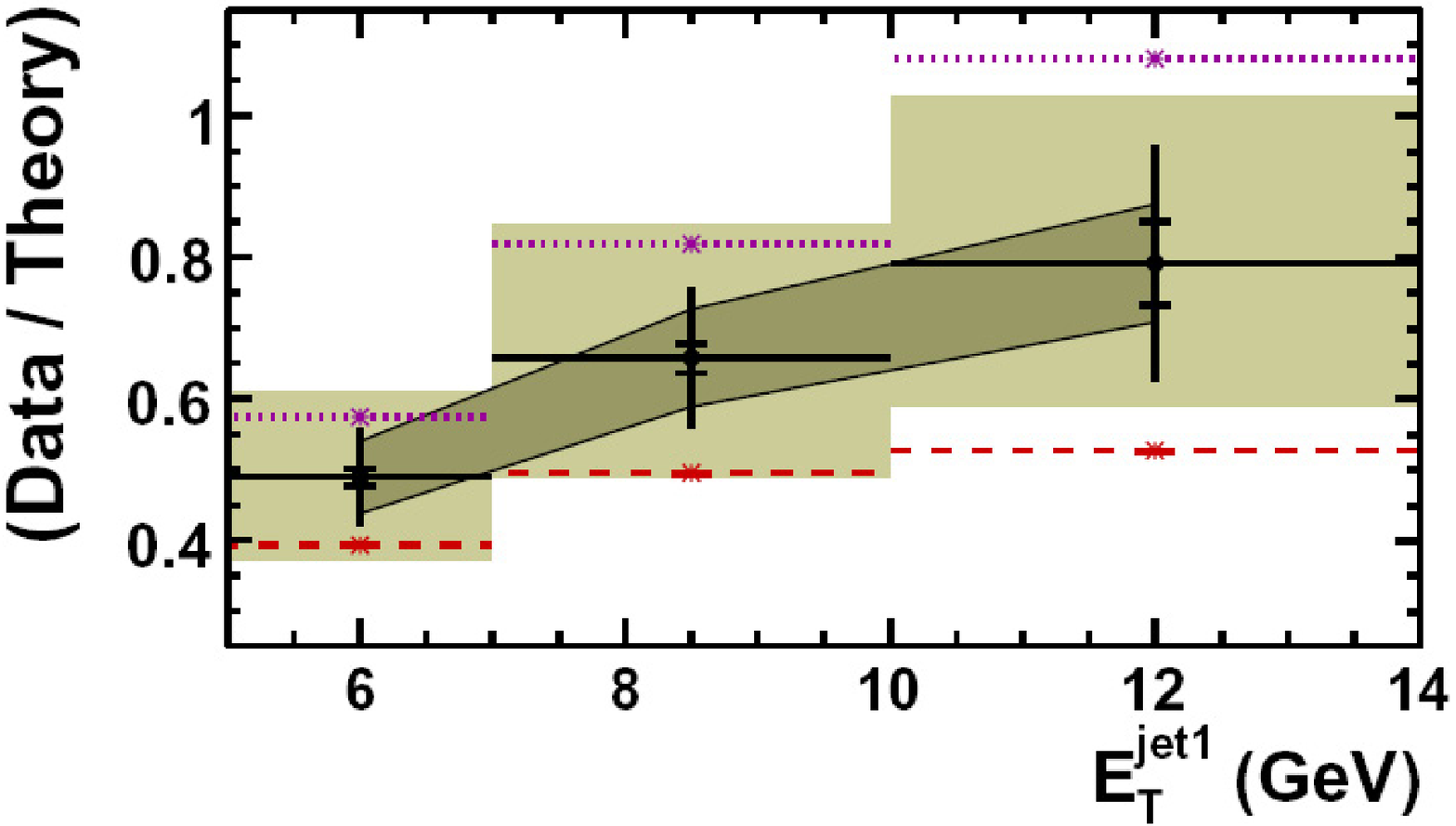}
\caption{Ratios of 
diffractive dijet photoproduction cross sections 
measured by H1 to NLO QCD calculations
\cite{karel}. 
}
\label{diffxsecs}
\end{wrapfigure}

As expected \cite{collins,fac:break},
DPDF-based predictions for hard
diffractive processes in $p\bar{p}$ scattering fail by around
an order of magnitude 
\cite{TevatronHERA}. This factorisation breaking is
generally attributed to absorptive corrections, corresponding to
the destruction of the outgoing
proton coherence and the rapidity gap due to
multiple interactions within a single event. These effects are associated
with the presence of a proton remnant, in contrast to the
point-like photon coupling in DDIS. The
corresponding `rapidity gap survival probability' can be
treated semi-quantitatively \cite{gap:survival} and its prediction
at LHC energies is a major current issue \cite{lhcdiff}. 

The questions of DPDF applicability and rapidity gap survival can be 
addressed
in hard diffractive photoproduction, where the virtuality of the
exchange photon coupling to the electron is close to 
zero \cite{butterworth:wing}.
Under these circumstances, the photon can develop
an effective partonic structure via $\gamma \rightarrow q\bar{q}$ fluctuations
and further subsequent splittings. In a simple leading order picture, there
are thus two classes of hard photoproduction: `resolved' 
interactions (figure~\ref{feynman}c), where
the photon interacts via its partonic structure and
only a fraction $x_\gamma$
of its four-momentum participates in the hard subprocess and
`direct' interactions (figure~\ref{feynman}b),
where the photon behaves as a point-like particle and
$x_\gamma = 1$.
The gap survival
probability has
been estimated 
to be 0.34 for resolved processes \cite{kkmr} and is
expected to be unity for direct photon interactions.

Figure~\ref{diffxsecs} \cite{karel}
shows ratios of H1 measurements of diffractive
dijet photoproduction cross sections to NLO QCD calculations which
neglect absorptive effects \cite{fr}. Results are shown
differentially in the leading jet transverse energy
$E_{T}^{jet1}$ and in hadron level estimators of
$z_{I\!\!P}$ and $x_\gamma$, obtained as described in \cite{sebastian}.
For most of the measured points, the ratios are significantly below
unity. When taking the H1 Fit B DPDFs \cite{h1:lrg}, 
which describe a wide range of
DDIS observables, there is little dependence
of the ratio on $z_{I\!\!P}$. 

\begin{wrapfigure}{r}{0.4\columnwidth}
\includegraphics[width=0.4\columnwidth]{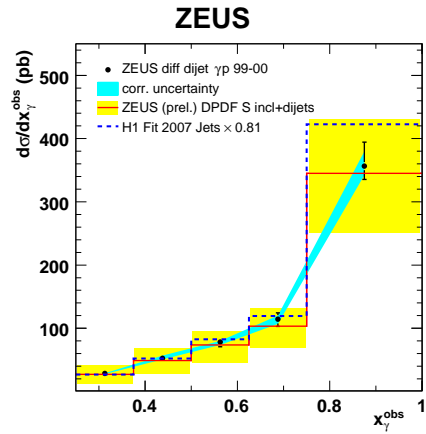}
\vspace*{-0.7cm}
\caption{ZEUS diffractive dijet photoproduction data \cite{zeus:gpjets},
compared with DPDF based predictions.}
\label{zeus:jets}
\end{wrapfigure}

Figure~\ref{zeus:jets} shows a recent ZEUS 
measurement \cite{zeus:gpjets} as a function of 
$x_\gamma$ compared with predictions based
on H1 \cite{matthias} and ZEUS \cite{zeus:diffqcd} DPDFs extracted
by fitting $\sigma_r^D$ and diffractive dijet electroproduction data. 
In contrast to the H1 case, these data are 
compatible with NLO predictions. A possible explanation for
the apparent discrepancy between the two collaborations is offered
by indications of a
dependence of the data-to-theory ratio on the jet transverse
energy 
(figure~\ref{diffxsecs}c)
\cite{sebastian,zeus:gpjets,karel}. 
The ZEUS measurement is made for $E_t^{\rm jet1} > 7.5 \ {\rm GeV}$,
whereas H1 measure for $E_t^{\rm jet1} > 5 \ {\rm GeV}$.
There is as yet no accepted theoretical explanation for this effect.

Intriguingly, the ratios of data to theory measured by both collaborations
have at most a weak dependence
on $x_{\gamma}$, in contrast to theoretical
expectations \cite{kkmr,klasen}.
Since the correlations between the variables are complicated
(e.g. $E_{T}^{jet1}$ and $x_\gamma$ are strongly positively correlated
through the kinematic restrictions),
more differential studies are required to fully
unfold the dynamics.

\section{Summary}

Recent H1 and ZEUS semi-inclusive diffractive DIS (DDIS) data are
in fair agreement within their normalisation 
uncertainties. 
The data exhibit proton vertex factorisation
to good approximation. Dependences on variables describing
the coupling to the proton lead to a picture in which DDIS probes
a diffractive exchange whose origins lie in the soft dynamics 
below typical factorisation scales, and which is similar to that
exchanged in soft hadronic 
scattering. The parton densities (DPDFs) associated with this
exchange
have a structure dominated by a hard gluon density,
which successfully describes all measured observables in diffractive
DIS, including the longitudinal diffractive structure function, $F_L^D$. 
The rapidity gap survival probability derived from 
DPDF-based predictions of hard diffractive 
photoproduction data is surprisingly similar for direct and resolved
photon interactions, a fact
which remains under investigation.


\begin{footnotesize}

\end{footnotesize}

\end{document}